\documentclass[aps,prl,twocolumn,superscriptaddress,groupedaddress]{revtex4}  % for review and submission
\usepackage{graphicx}  % needed for figures
\usepackage{dcolumn}   % needed for some tables
\usepackage{bm}        % for math
\usepackage{amssymb}   % for math
\usepackage{color}     % for font colors

\begin{document}

% The following information is for internal review, please remove them for submission
%\widetext
%\leftline{Version xx as of \today}
%\leftline{Primary authors: Satoshi Sugimoto}
%\leftline{To be submitted to PRL}
%\leftline{Comment to {\tt d0-run2eb-nnn@fnal.gov} by xxx, yyy}
%\centerline{\em D\O\ INTERNAL DOCUMENT -- NOT FOR PUBLIC DISTRIBUTION}

% the following line is for submission, including submission to the arXiv!!
%\hspace{5.2in} \mbox{Fermilab-Pub-04/xxx-E}

\title{Dynamics of coupled vortices in a pair of ferromagnetic disks}

\author{Satoshi~Sugimoto}
\affiliation{Institute for Solid State Physics, University of Tokyo, 5-1-5 Kashiwa-no-ha, Kashiwa, Chiba 277-8581, Japan}
\author{Yasuhiro~Fukuma}
\affiliation{Advanced Science Institute, RIKEN, 2-1 Hirosawa, Wako, Saitama 351-0198, Japan}
\author{Shinya~Kasai}
\affiliation{National Institute for Material Sciences, Sengen, Tsukuba 1-2-1, Japan}
\author{Takashi~Kimura}
\altaffiliation{Present address: Inamori Frontier Research Center, Kyushu University, 744 Motooka, Nishi-ku, Fukuoka 819-0395, Japan}
\affiliation{Institute for Solid State Physics, University of Tokyo, 5-1-5 Kashiwa-no-ha, Kashiwa, Chiba 277-8581, Japan}
\author{Anjan~Barman}
\affiliation{Department of Material Sciences, S. N. Bose National Centre for Basic Science, Block JD, Sector III, Salt Lake, Kolkata 700 098, India}
\author{Y. Otani}
\email{yotani@issp.u-tokyo.ac.jp}
\affiliation{Institute for Solid State Physics, University of Tokyo, 5-1-5 Kashiwa-no-ha, Kashiwa, Chiba 277-8581, Japan}
\affiliation{Advanced Science Institute, RIKEN, 2-1 Hirosawa, Wako, Saitama 351-0198, Japan}
\date{\today}

\begin{abstract}
We here experimentally demonstrate that gyration modes of coupled vortices 
can be resonantly excited primarily by the ac current in a pair of ferromagnetic disks 
with variable separation. 
The sole gyration mode clearly splits into higher and lower frequency modes via dipolar interaction, 
where the main mode splitting is due to a chirality sensitive phase difference in gyrations of the coupled vortices, 
whereas the magnitude of the splitting is determined by their polarity configuration. 
These experimental results show that the coupled pair of vortices behaves similar to a diatomic molecule 
with bonding and anti-bonding states, implying a possibility for designing the magnonic band structure 
in a chain or an array of magnetic vortex oscillators. 
\end{abstract}

%\pacs{72.25.Ba, 72.25.Mk, 75.70.Cn, 75.75.-c}
\maketitle

%\section{\label{sec:level1}First-level heading}
% sections are not used for PRL papers

Magnetic vortex structure \cite{1,2} is one of the fundamental spin structures 
observed in submicron-sized ferromagnetic elements.
It is well characterized by two degrees of freedom,
one is 'chirality' ($c = \pm 1$),  direction of the in-plane curling magnetization along the disk circumstance,
and the other is 'polarity' ($p = \pm 1$), direction of the out-of-plane core magnetization.
Particularly in the case of the disk, only the core region whose typical size is 
$\sim$ 10 nm generates the stray field, the static interaction between vortices in an array 
is thus negligibly small in the ground state. However in the low frequency excitation state 
called the gyration (or translational) mode \cite{3,4},
the surface magnetic charges appear with the core motion, 
that brings about the dynamic dipolar interaction between vortices.

Coupled pair of magnetic vortices can be considered as a vortex molecule bound via dipolar interaction, 
which is a mimic of diatomic molecule with the van der Waals bonding \cite{5}.
The bonding or anti-bonding state respectively corresponds to in-phase or out-of-phase 
gyration of coupled vortices whose detailed energy levels are decided by combination of chiralities and polarities. 
This dynamic coupling is also effective in a two-dimensional array system \cite{6,7},
and thus allows us to design the density of states of the eigenfrequencies, 
the so-called "magnonic band structure", 
by arranging the core polarizations in a two-dimensional array. At the moment, 
there are few experimental reports for the coupled vortices via direct exchange interaction \cite{8,9} 
and also dipolar interaction in physically separated vortices \cite{10,11,12,13}.
However the problem is still wide open in terms of experimental determination of the detailed condition 
for the mode splitting and the magnitude of the dipolar coupling in interacting vortices.

Herein we demonstrate the experimental evidence of the resonant excitation of a magnetostatically coupled pair 
of vortices as a clear mode splitting of spectra. The partial excitation using ac current 
causes the energy transfer via dipolar interaction between two vortices, 
which results in a collective excitation of coupled gyrations. 
The observed mode splitting is reproduced by both micromagnetic simulation and analytical calculation. 
The different combinations of core polarities and phase difference in gyrations give rise to four distinct modes. 
The magnitude of the vortex coupling is also discussed.

\begin{figure}[here]
\includegraphics[scale=0.42]{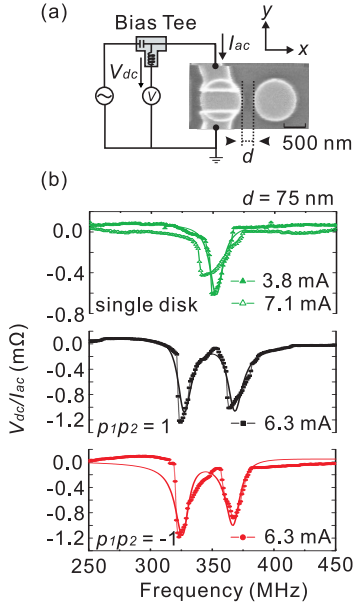}
\caption{\label{fig:epsart} (color on line). (a) Schematic diagram of the measurement circuit and an SEM image of the sample. 
Two Copper electrodes are attached to one of the disks in the Permalloy disk pair and the enveloped core is excited by a radio frequency current. 
Dynamics of the cores can be detected as dc voltages through the spin-torque diode effect utilized by a resistance oscillation associated with the core gyration.
A lock-in technique is adopted at room temperature.
(b) Frequency dependence of the normalized dc voltage $V_{dc}/I_{ac}$ 
measured for an isolated disk (green triangles) 
and for the paired disks with different polarities; black squares for $p_1p_2 = 1$, 
and red circles for $p_1p_2 = -1$. 
The ac current amplitudes $I_{ac}$ used for the measurements are $I_{ac} = 3.8$ mA and 7.1 mA for the single disk, 
and $I_{ac}=6.3$ mA for the paired disks with the edge to edge distance $d$ = 75 nm.
Solid curves in each spectrum represents the best fit to the data points using Eq. (1), 
thereby the dipolar coupling is evaluated.}
\end{figure}

Figure 1(a) shows a scanning electron microscope (SEM) image of the sample together with the measurement circuit.
Each disk has the same dimension, 500 nm in disk radius r and 50 nm in thickness. 
Two neighboring Permalloy (Fe$_{19}$Ni$_{81}$; Py) disks with variable edge to edge separation $d$ from 75 nm to 250 nm 
and electrical leads are fabricated on thermally oxidized Si (100) by means of electron beam lithography combined 
with electron beam evaporation techniques.
The polarities of vortices were confirmed by means of magnetic force microscopy prior to all the electrical measurements. 
Hereafter, the number 1 or 2 respectively represents the vortex confined in the left or that in the right disk 
shown in the SEM image of Fig. 1.
% add in ref
One of the paired vortices is excited by a radio frequency ac current $I_{ac}$ \cite{14,15,16,a,b} 
and the resulting dc voltage $V_{dc}$ through a bias tee is synchronously detected by the same electrical contact probes \cite{15,18}. 
In order to set the configuration of polarities $p_1p_2 = 1$ or $p_1p_2 = -1$, 
$p_1$ of the excited vortex is switched by applying high $I_{ac}$ of about 20 mA ($3.5 \times 10^{11}$ A/m$^2$) 
at the resonant frequency \cite{19}.

Figure 1(b) shows the measured $V_{dc}/I_{ac}$ as a function of the frequency of $I_{ac}$ for the Py disk pair 
with $d$ = 75 nm. 
The reference spectrum for single disk shown by green symbols exhibits a sole dip at 352 MHz, 
which corresponds to the resonant frequency of the vortex core gyration. 
When $I_{ac}$ is applied to one of the two neighboring Py disks, 
clear mode splitting takes place as can be seen in the black and red spectra in the figure. 
Important to note is that the gyration mode of the single vortex with $I_{ac}= 7.1$ mA only shows lower frequency shift 
and asymmetric broadening due to nonlinear effects \cite{20} as plotted by open symbols. 
Therefore, the observed mode splitting is primarily due to the effect of the dipolar coupling between the disks 
mediated by magnetic side charges. 
The magnitude of the mode splitting for $p_1p_2 = -1$ is slightly enhanced by several MHz compared to that for $p_1p_2 = 1$. 

\begin{figure}[here]
\includegraphics[scale=0.42]{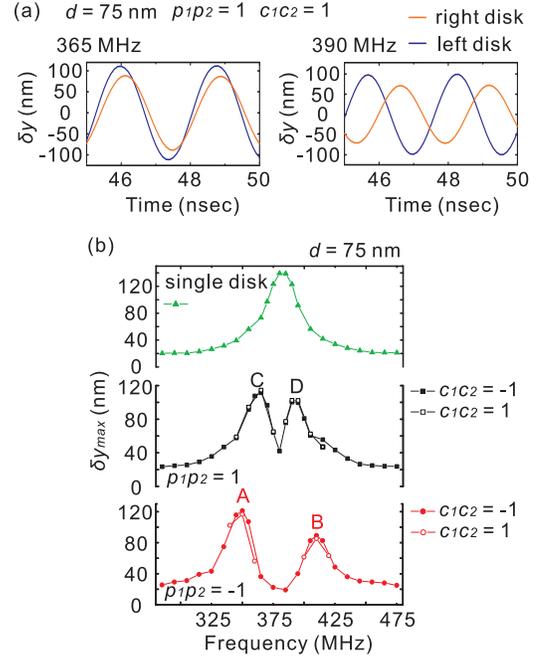}%Here is how to import EPS art
\caption{\label{fig:wide}(color on line). (a) Simulated time evolutions of vortex cores 
at resonance frequencies for ($p_1p_2$, $c_1c_2$) = (1,1) under ac currents (365 MHz and 390 MHz). 
Blue solid lines show the motions of the current-excited core in the left disk and 
red lines correspond to those of the indirectly excited core in the right disk.
(b) Dispersion relations of amplitude of steady gyrations. 
Values of $\delta y_{max}$ show the radii of steady gyrations 
(50 nanoseconds after beginning of the current flow). 
Black symbols correspond to the parallel polarities $p_1p_2 = 1$ and 
red ones to anti-parallel polarities $p_1p_2 = -1$ for opposite chiralities $c_1c_2 = -1$. 
Results of same chiralities $c_1c_2 = 1$ are plotted by open symbols. 
Simulation results for a single vortex are also presented by green symbols for comparison.} 
\end{figure}

To gain insights into the dynamics of magnetostatically coupled vortices, 
micromagnetic simulations based on the Landau-Lifshitz-Gilbert equation \cite{21} 
were performed on pair of Py disks with identical physical dimensions. 
Typical material parameters for Py are used : the saturation magnetization $M_s = 1$ T, 
the exchange stiffness constant $A = 1.05 \times 10^{-11}$ J/m, 
the spin polarization $P$ = 0.4 and the damping coefficient $\alpha = 0.01$. 
The disk is divided into rectangular prisms of $5 \times 5 \times 50$ nm$^3$ for the simulation. 
A uniform $I_{ac}$ of $1.2 \times 10^{11}$ A/m$^2$ is applied only to the left disk.

After several nanoseconds from the start of excitation, 
the core gyration settles in an almost circular orbit and its amplitude is strongly enhanced at the resonance frequency \cite{14,15}.
This induces the core gyration in the neighboring disk, 
which also settles in the steady circular orbit, 
and the collective gyration of the two vortices becomes fully synchronized and 
the eigenfrequencies of these modes appear in the spectra as characteristic resonance frequencies. 
Figure 2(a) represents the time evolution of the core deviations $\delta y$ at lower and higher resonance frequencies 
with respect to the single vortex for parallel polarities $p_1p_2 = 1$ and the same chiralities $c_1c_2 = 1$. 
At the lower frequency (365 MHz) both left and right cores rotate almost in-phase, 
whereas at the high frequency (390 MHz) the phase of the right core is retarded by approximately half a period. 
This is very much in analogy with covalent bonding in diatomic molecules or other forms of coupled oscillators. 
In the case of in-phase excitation, 
magnetic charges at side surfaces form magnetic dipoles resulting in attractive force between two cores, 
which corresponds to a bonding orbital. 
On the other hand, the side charges of the disks repel each other for the out-of-phase excitation, 
having an analogy with an anti-bonding orbital.

Figure 2(b) shows the maximum deviation of the core from the center ($\delta y_{max}$) 
of the steady orbital for the current excited vortex in the left disk 
as a function of the ac frequency for the sample with $d$ = 75 nm. 
The single vortex has a clear resonance peak at around 380 MHz and exhibits a small discrepancy with the experimental result 
(Fig. 1(b)) caused by self reduced magnetization \cite{19}.
For the coupled vortices, 
two clear resonance peaks are observed on both higher and lower frequencies relative to the sole peak for the single vortex. 
The magnitude of mode splitting for $p_1p_2 = -1$ is larger than that for $p_1p_2 = 1$, 
as experimentally observed in Fig. 1 (b). 
At both low and high frequencies, the in-phase and out-of-phase modes are degenerated with respect to the chirality. 
Figure 3 summarizes all four resonance modes from A to D characterized by rotational directions and 
the phase difference in gyrations. 
The lower frequency modes (A and C) stabilize with the help of attractive interaction due to the magnetic charges 
appearing along the disk circumferences. 
On the other hand, the higher frequency modes (B and D) stabilize with the help of repulsive interaction due to 
the opposite magnetic charges. 
Therefore the appearance of either in-phase or out-of-phase gyration in Fig. 2 (a) depends only on 
the polarity configuration $p_1p_2$.

\begin{figure}[here]
\includegraphics[scale=0.42]{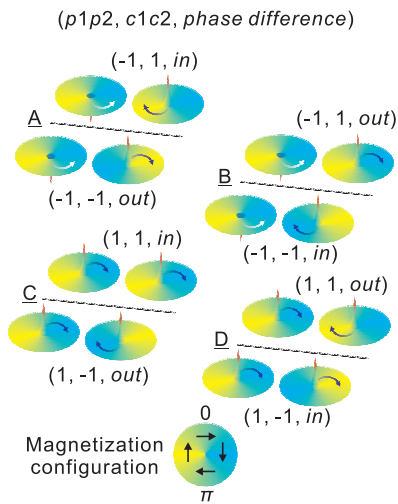}%Here is how to import EPS art
\caption{\label{fig:wide}(color on line). Schematic diagrams of four different resonance modes. 
Each mode is identified with the sign of polarities $p_1p_2$, chirarities $c_1c_2$ and the phase difference of the gyrations.}
\end{figure}

\begin{figure}[here]
\includegraphics[scale=0.42]{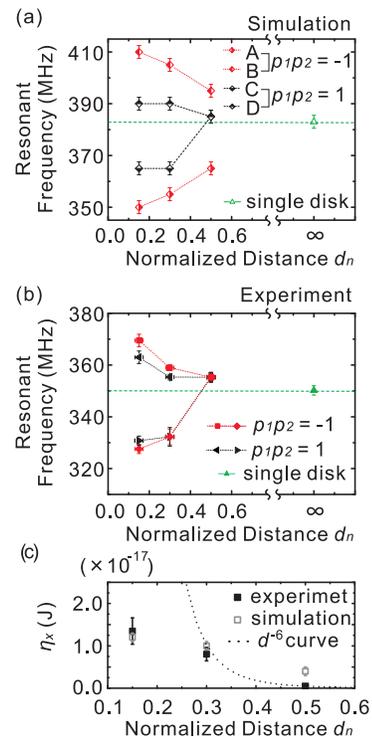}
\caption{\label{fig:epsart} (color on line). Eigenfrequencies of the coupled gyration modes as a function of the separation distances. 
The simulation and experimental results are shown in (a) and (b), respectively. 
The separation distance is given by a dimensionless value $d_n = d/R$.
(c) Estimated values of coupling strength $\eta_{x}$ from both experiment and simulation with different separation distances.
A $d^{-6}$ curve from the rigid vortex model is also plotted as a dotted line.}
\end{figure}

In both experimental and simulated results, 
the splitting amplitude tends to be enhanced with the decrease in the normalized separation distance $d_n = d/r$, 
as shown in Fig. 4. 
To check the effect of the current induced Oersted field on the excitation of the gyration of the right Py disk, 
the simulation was performed by replacing the left Py disk with an electrode where the current flows with the identical condition. 
The excited gyration amplitude due to the Oersted field is much smaller than that for the coupled vortices, 
implying that the magnetic dipolar interaction is the dominant factor for the indirect excitation of the right Py disk.

We here discuss the experimental results analytically on the basis of Thiele's equation \cite{22,23}, 
which describes the gyrations of vortex cores $r_i = (X_i, Y_i)$ with $i = 1, 2$. 
%add
%start
%\textcolor{red}{
It should be noted that our calculations take into account only the adiabatic spin transfer torque
as the excitation force of the vortex 1 in the left disk to simplify the analyses.
The non-adiabatic and the Oersted field terms for the excitation are neglected.
%add below sentence about beta
The effect of the non-adiabatic term fundamentally appears as the phase and
the rotational radius shifts of the gyroscopic motion of the vortex core.
%described as $\delta \phi = \tan^{-1}{\beta D/G}$ and $\delta r = \sqrt{G^2 + (\beta D)^2}$ using
%gyrotropic constant $G$, damping tensor $D$ and non-adiabatic coefficient $\beta$.
%Even if the $\beta$ is large ($\sim 0.1$) \cite{17}, only the spectrum intensity increases by about 10 \%
%in Fig. 2.
%add sentence about Oersted field
One would point out the current induced Oersted field as the other origin of core excitation.
Both the terms would affect the dynamics of the vortex 1 \cite{16,a,b,17},
however, may not influence the coupled modes identified in Fig. 3.
%}
%end
Herein, the detected dc voltage can be evaluated by the following equation \cite{18};
\begin{equation}
V_{dc} = I_{ac}C/2 (\textrm{Re}[X_1]\cos{\delta_1}+\textrm{Re}[Y_1]\sin{\delta_1}),
\end{equation}
where $C$ is a constant and $\delta_1$ is the phase difference between core and resistance oscillations in the left disk.
The model \cite{5} assumes the magnetostatic interaction energy term in Thiele's equation as
\begin{equation}
U_{int} = c_1c_2/R^2(\eta_{x}X_1X_2 - \eta_{y}Y_1Y_2)+O(|r/R^3|).
\end{equation}
According to the rigid vortex model \cite{22}, 
the values of $\eta_{x}$ and $\eta_{y}$ are decided only by the shape of ferromagnetic element and the separation distance, 
independent from the excitation amplitude.
%remove
%One should note that this analytical model does not accurately describe the actual spin dynamics in terms of several aspects;
%neglect of the nonlinear effect about core gyrations, current induced Oersted field 
%and non-adiabatic spin transfer torque \cite{16,a,b,17} in addition to 
%the overestimate of the stray field in the rigid vortex model \cite{24}. 
%end
As a simple solution for these problems, 
$\eta_{x}$ and $\eta_{y}$ should be treated as phenomenological fitting parameters including the influence of volume charges under the assumption of 
complete linear response with the fixed parabolic potential 
$U = 1/2 k_x X_i^2 + 1/2 k_y Y_i^2 ~(k_x, k_y \equiv const.)$ 
and the interaction energy $U_{int}$. 
The best fitted curves to the experimental data using the above equation are shown in Fig. 1(b) by solid curves. 
The values of $k_x$ and $k_y$ are decided by a result of single vortex with $I_{ac} =$ 3.8 mA as 
($k_x$, $k_y$) = ($4.5 \times 10^{-4}$ J/m$^2$, $6.4 \times 10^{-4}$ J/m$^2$).
These results agree well with the previous work \cite{18}, 
implying adequacy of Thiele's equation for the present analyses.
The obtained values from experiments with 75 nm separation are 
($\eta_{x}$, $\eta_{y}$) = ($1.4 \times 10^{-17}$ J, $4.9 \times 10^{-22}$ J) 
and those from simulations performed under identical conditions are 
($\eta_{x}$, $\eta_{y}$) = ($1.2 \times 10^{-17}$J, $8.1 \times 10^{-18}$J).
While the value of $\eta_x$ agrees well each other, a large discrepancy of $\eta_y$ is observed. 
This is possibly due to anisotropy along $y$-direction in the experiment 
due to the presence of attached Cu electrodes, 
which causes deviation of stray field from the simulation, 
resulting in closer resonance frequencies for coupled vortices with $p_1p_2 = \pm 1$ 
(red and black symbols) in Fig. 4(b). 
The separation dependence of $\eta_x$ is shown in Fig. 4 (c). 
It is clear that the coupling strength increases　with　the decrease in　the　separation　distance,
which can　be　clearly　reproduced by the two-dimensional micromagnetic simulation. 
%The results reveal that the contributions due to the other current effects, 
%such as the nonadiabatic spin torque \cite{17}
%and the Oersted field induced by current distributions is relatively small compared to the spin torque. 
It should be noted that the $d$ dependence of $\eta_x$ does not follow the $d^{-6}$ dependence expected from a rigid vortex model \cite{5,6}. 
%\textcolor{red}{
A strong magnetostatic coupling modifies the trajectories of the gyrations and the magnetization configuration near the edge region 
thus causes a deviation from the model, where a circular magnetization around the core is assumed.
%}

In summary, 
we have experimentally demonstrated the resonant excitation of coupled gyration modes in paired vortices 
by means of local excitation by an ac current passing through one of the disks in the pair. 
Excited coupled modes are identified by rotational directions and a phase difference as four different eigen-modes. 
The unique property in this system give us a guiding principle for designing the magnonic crystal 
in further expanded systems such as one-dimensional chains and two-dimensional arrays and 
a candidate for novel tunable oscillators using vortices \cite{6,9,25}.

%ここまで
%to add Prof nakatani
We would like to thank Y.~Nakatani for fruitful discussions.
This work is supported in part by a 
Giant-in-Aid for Scientific Research in Priority Area 
"Creation and Control of Spin Current" (Grant No. 19048013) 
from the Ministry of Education, Culture, Sports, Science and Technology of Japan.

\end{document}